\documentclass[12pt,twocolumn]{iopart}
\usepackage{graphicx,gensymb,subfigure}

\usepackage{bm,cite,amssymb,color}

\begin{document}

\title{Plastic Deformation of 2D Crumpled Wires}

\author{M A F Gomes$^{1}$\footnote{Corresponding author e-mail: mafg@ufpe.br; fax: ++55 81
3271 0359;  phone: ++55 81 2126 7614}, V P Brito$^{2}$, A S O Coelho$^{2}$, and C C Donato$^{1,3}$}

\address{$^1$Departamento de F\'{\i}sica, Universidade Federal de
Pernambuco, 50670-901, Recife - PE - Brazil}

\address{$^2$Departamento de F\'{\i}sica, Universidade Federal do Piau\'{\i}, 64049-550, Teresina - PI - Brazil}

\address{$^3$Centro Internacional de F\'{\i}sica da Mat\'eria Condensada, Instituto de F\'{\i}sica, Universidade de Bras\'{\i}lia, 70910-900, Bras\'{\i}lia - DF - Brazil}

\begin{abstract}

\indent When a single long piece of elastic wire is injected trough channels into a confining two-dimensional cavity, a complex structure of hierarchical loops is formed. In the limit of maximum packing density, these structures are described by several scaling laws. In this paper it is investigated this packing process but using plastic wires which give origin to completely irreversible structures of different morphology. In particular, it is studied experimentally the plastic deformation from circular to oblate configurations of crumpled wires, obtained by the application of an axial strain. Among other things, it is shown that in spite of plasticity, irreversibility, and very large deformations, scaling is still observed.

\end{abstract}


\maketitle

\section{\label{sec1} Introduction}

\begin{figure*}[!]
\centering
\includegraphics{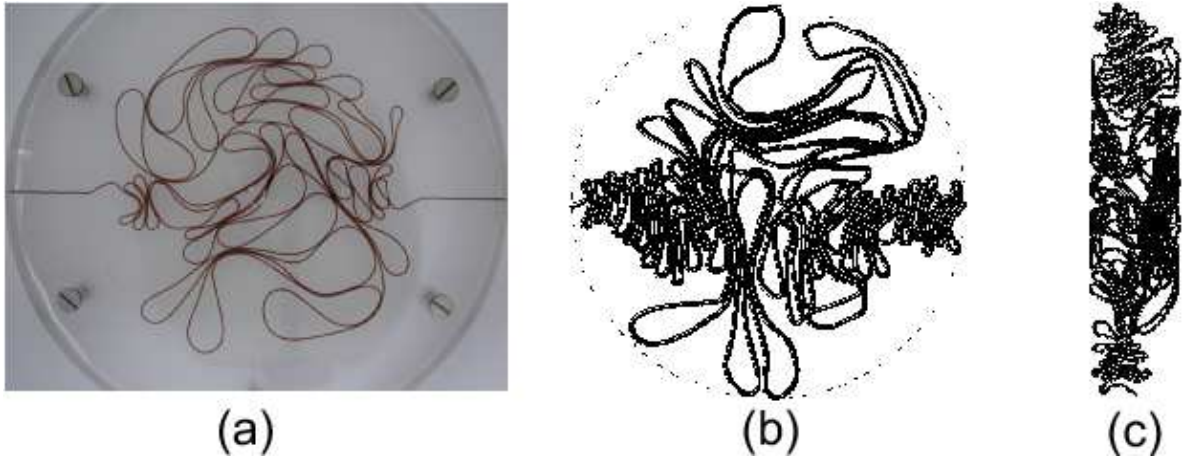}
\caption{\label{fig1}Photographs of rigid heterogeneous hierarchical structures of loops obtained from forced injection of wire of diameter $\zeta$ in a quasi-two-dimensional cavity of height $h$, and radius $R$: (a) copper wire, $\zeta = 1$~mm, $h=1.1$~mm, and $R=100$~mm; (b) $\mbox{Pb}_{0.40}\mbox{Sn}_{0.60}$  wire, $\zeta=1.5$~mm, $h=1.6$~mm, and $R=75$~mm. (c) Photograph of the sample shown in~\ref{fig1}(b) after axial deformation of $69\%$. }

\end{figure*}

\indent In the last two decades, crumpled surfaces have been a subject of increasing interest in theoretical and applied physics in connection with fractals, mechanical properties, nontrivial scaling laws, packing processes, anomalous relaxation, condensation of stress, among other aspects~\cite{1,2,3,4,5,6,7,8,9,10,11,12,13,14}. On the other hand, heterogeneous structures of crumpled wires, with a quasi-one-dimensional topology, were much less studied. However, it is known that ill-defined compression procedures of wires in three-dimensional space give origin to robust scaling laws for this type of disordered system~\cite{15}. Only recently rigid, non-compact, heterogeneous crumpled structures of wires of circular shape were obtained in two dimensions (shortly, 2D-CW) from packing processes of a single layer of wire injected trough channels into a quasi-two-dimensional cavity~\cite{16,17,18,19,20,21,22}. In particular, in the first studies of 2D crumpled structures were used wires of copper or steel and nylon fishing lines~\cite{16, 17, 19}, which give origin to quasi-reversible or almost perfectly reversible packing structures of loops of the type illustrated in figure~\ref{fig1}(a). The degree of reversibility mentioned here can be evaluated for instance, from the level of recovery or uncoiling of the wire when the confining wall of the cavity is withdrawn. Thus, considerable elastic energy remains stored in the cavity in different degrees after a long period if wires of these materials are used to generate the crumpled structures. The low dimensional packing of an elastic wire in a quasi-two-dimensional cavity of circular or square shape generates complex configurations of a cascade of loops with a fractal dimension $D = 1.9 \pm 0.1$ as obtained from box-counting and mass-radius measurements~\cite{16}. Although the mass of wire in these heterogeneous packing structures distributes in an essentially two-dimensional support, from the practical point of view the maximum packing fraction observed within the cavity is significantly less than the unit. In fact, it is close to $0.15$ (i.e. much smaller than typical packing fractions of $0.82-0.84$ obtained with the random close packing of discs), irrespective the material of the wire, the angle formed between the injection channels, and other details~\cite{16, 17, 19}.

\indent Differently, in the present work use is made of plastic wires of the alloy $\mbox{Pb}_{0.40}\mbox{Sn}_{0.60}$, which in turn give origin to strongly irreversible structures of 2D-CW of the type exemplified in figure~\ref{fig1}(b). When compared with the elastic case, crumpled structures of a plastic wire present both a higher packing fraction (in our experiments, it is close of $0.30$), and a different morphology for the individual loops, as well as for the heterogeneous distribution of the loops, as will be detailed in the next section. Besides its intrinsic interest, low-dimensional structures of crumpled wires can, in principle, be related with important topics in applied physics as for instance, self-avoiding walks and polymer configurations on the plane~\cite{23}, or with random network motifs of atoms or molecules adsorbed on solid surfaces~\cite{24}. Additionally, crumpled wires could be of interest in the study of the packing of genetic material in chromosomes and viral capsids~\cite{21}, and to describe the statistical aspects of meandering nanotubes in a planar confining geometry~\cite{25}.

\indent In this paper, using experiments and scaling arguments it is studied the plastic deformation of the nearly circular configurations of 2D-CW (figure~\ref{fig1}(b)) into oblate configurations (figure~\ref{fig1}(c)), obtained as a consequence of the application of an axial strain. The morphologies shown in figure~\ref{fig1}(b) and~\ref{fig1}(c) rigorously present a spontaneous break of the rotational and translational symmetries of the free space, albeit present scaling symmetry in several aspects. Figure~\ref{fig1}(c), in particular, represents a structure with less porosity halfway towards a more condensate state akin to a disordered solid.

\indent The structure of the paper is the following: in section~\ref{sec2} the experiments of deformation of 2D-CW are described in detail, in section~\ref{sec3} the experimental results are presented and discussed, and in section~\ref{sec4} there is a summary of the main results.

\section{\label{sec2} Experimental Details}

\begin{figure*}[!]
\centering
\includegraphics{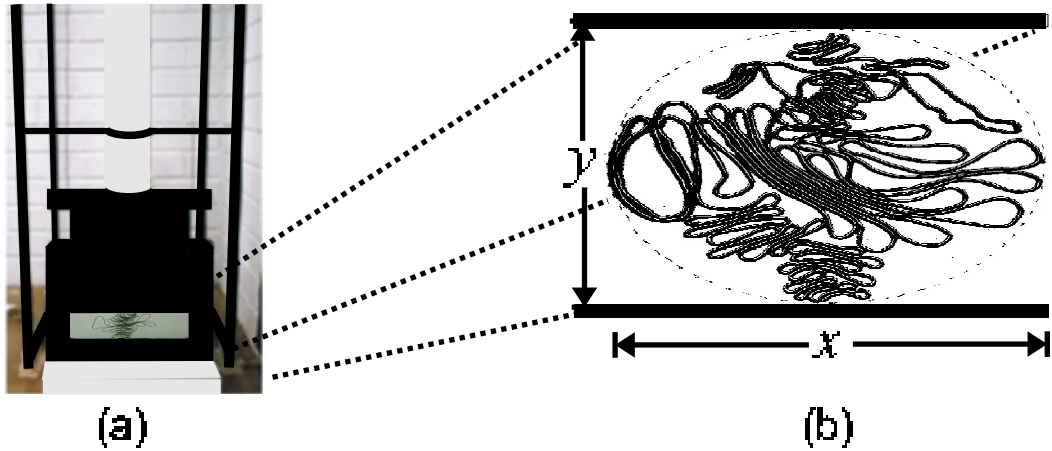}
\caption{\label{fig2}(a) Experimental setup used to deform axially and irreversibly the crushed wires of the alloy $\mbox{Pb}_{0.40}\mbox{Sn}_{0.60}$  of the type shown in figure~\ref{fig1}(b). (b) The main physical variables measured in the experiment: the height $y$ and the equatorial diameter $x$ of the system.}

\end{figure*}

\indent In figure~\ref{fig2}(a) it is shown the experimental setup used to deform axially and irreversibly the crushed wires of the alloy $\mbox{Pb}_{0.40}\mbox{Sn}_{0.60}$. This wire which is commonly used in welding in electronic devices had a diameter $\zeta = 1.5$~mm, and the internal circular cavity of the cell where the initial configurations of crushed wires were generated (not shown) had a radius $R = 75$~mm. Figure~\ref{fig2}(b) illustrates the main physical variables measured in the experiment. From the measurements of height $y$ and the equatorial diameter $x$ one can obtain both the vertical strain, $\delta_y = (y-2R)/2R$, and the transversal strain, $\delta_x = (x-2R )/2R$.

\indent Several configurations of 2D-CW were used in this work, each one at the maximum packing density (figure~\ref{fig1}(b)): the average length $L$ of plastic wire introduced in the cavity was $3700 \pm 370$~mm, corresponding to a packing density $p = \zeta L/\pi R^2 = 0.3140 \pm 0.0314$, and to an average aspect ratio $\zeta/L = 4.05 \times 10^{-4}$ for the wire.

\indent Each injection experiment begins fitting a straight wire in the opposite channels and subsequently pushing manually and uniformly the wire on both sides of the cell toward the interior of the cavity. Here all 2D-CW configurations were obtained with the regime of injection along opposite channels ($180^\circ$ angular aperture). The injection velocity at each channel in these experiments was of the order of $1$~cm/s. For wires with the largest lengths, the crumpled structures become rigid, the difficulty in the injection increases rapidly, and the injection velocity goes to zero with the formation of a jammed state of crumpled wire within the cavity. However, the observed phenomena are widely independent of the injection speed for all interval of injection velocity compatible with a manual process. At this stage the circular cell is maintained in the horizontal position. The interested reader is referred to previous works~\cite{16,17,19} for further details on how to obtain the nearly circular 2D-CW. After the jammed stage to be reached, the 2D-CW configurations are transferred to the rectangular cell which is disposed vertically in the press frame as indicated in figure~\ref{fig2}(a).

\indent The press shown in figure~\ref{fig2}(a) has a transparent vertical cell formed from two rectangular pieces of glass $1.2$~cm thick parallely disposed and separated by a distance of $1.5$~mm allowing only a single layer of crumpled wire. The cavity of the cell was polished in order to reduce the friction. Cavity and wire operated in dry regime, free of any lubricant. The press has a rigid metallic piece that fits with precision the separation between the glass plates and it is used to impose the strain on each sample along the direction of the original axis of injection. The circular configurations were subsequently and progressively submitted to vertical strains $\delta_y$ in the interval $- 0.69 \leq \delta_y \leq 0$, in  steps of 0.033. The forces necessary to reach the largest deformations are close to $600$~N. With the introduction of the strain along the $y$-axis, the 2D-CW structures break their approximate globular symmetry and become progressively oblate as shown in figure~\ref{fig1}(c), which illustrates the type of observed configuration for vertical strain $\delta_y = - 0.69$, corresponding to the initial structure shown in figure~\ref{fig1}(b).

\indent At this point, it is interesting to compare some basic and qualitatively new aspects
of the irreversible 2D-CW exemplified in figure~\ref{fig1}(b), with the more elastic and less irreversible structures of crumpled wires previously studied (figure~\ref{fig1}(a))~\cite{16, 17, 19}. Firstly, the plastic structures are more heterogeneous and a scaling cascade of loops is much less evident. Secondly, the plastic structure present more loose domains, with very low density of wire-wire contacts. However, and in spite of these low density domains, the global packing fractions of the plastic configurations of CW are larger than in the less irreversible case. This is due to the fact that elastic wires resist more efficiently to confinement. Thirdly, there is a tendency for the wire to distribute itself in some regions of the circular cavity adopting columnar structures as a stack of lamellae, with some orientational order, and reminiscent of the structure of smectic liquid crystals~\cite{26}.

\section{\label{sec3} Results and Discussion}

\begin{figure*}[!]

\subfigure[]{\includegraphics{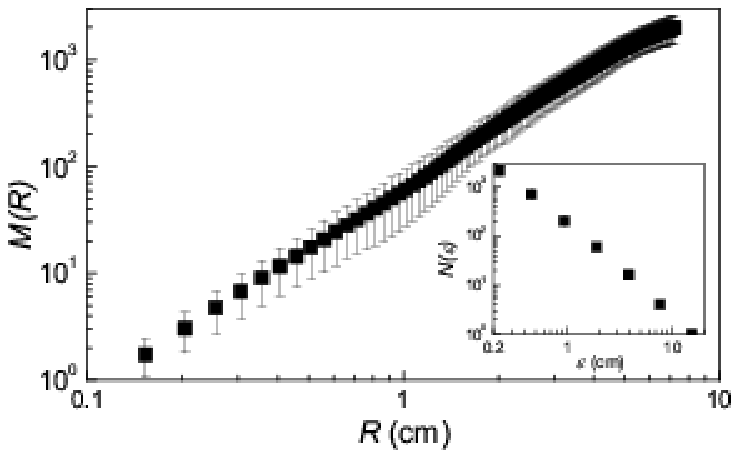}}
\subfigure[]{\includegraphics{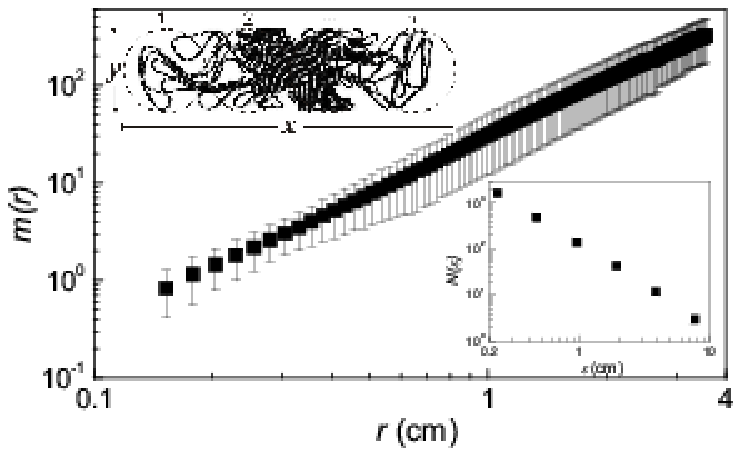}}
\caption{\label{fig3}(a) Main plot: mass ($M$)-size ($R$) dependence for all samples of undeformed 2D-CW studied in this paper at the maximum packing density (figure~\ref{fig1}(b)). The power law fit for the ensemble, $M \sim R^D$, has $D=1.93\pm0.12$. The scaling region covers typically the interval $0.15 < R(\mbox{cm}) < 5.0$. The inset shows the corresponding box-counting plot $N(\epsilon)$ leading to the same exponent $D$ which is obtained from the mass-size scaling. (b) mass ($m$) - size ($r$) data after averaging in all blobs associated with the structures of the type shown in figure~\ref{fig1}(c) (mean vertical deformation $\left\langle |\delta_y| \right\rangle = 0.65 \pm 0.04$). The best fit obtained indicates that $m \sim r^D$, with $D = 1.90 \pm 0.10$. The upper left inset shows a typical distribution of blobs for 2D-CW at high strain. The lower right inset exhibits the corresponding box-counting plot for 2D-CW at high strain leading to the same exponent $D$ previously reported.}

\end{figure*}

\indent Figure~\ref{fig3}(a) shows in the main plot the average mass ($M$) - size ($R$) dependence for the ensemble of 2D-CW studied in this paper when the jammed state is reached (figure~\ref{fig1}(b)). The inset shows the corresponding box-counting plot~\cite{16,19}. Both figures are compatible with a fractal dimension $D = 1.93 \pm 0.12$, as previously observed by using much less irreversible crushed structures made with wires of copper and steel, as well as with nylon fishing lines~\cite{16, 19}. Interestingly, the fractal mass-size dimension seems insensitive to the elastic properties of the wire, and to the degree of irreversibility involved. This supports the idea that in crumpled systems, $D$ is heavily dependent on the self-avoidance forces, and on the fixed quasi-one-dimensional topology of the wire~\cite{14}. In the context of the present discussion, it is interesting to mention that it was recently noticed that, in nature, DNA in many bacteriophage viral capsids present a similar mass-size scaling relation in the form $\mbox{DNA-length} \sim \mbox{DNA-mass} \sim \mbox{(capsid-size)}^{1.9}$~\cite{21}.  Of course, this is a surprising result because a viral capsid is a three-dimensional cavity and, in principle, we should expect a mass-size exponent close to 3. Furthermore, it is observed the striking independence of the fractal mass-size exponent with the degree of repacking of the wires, i.e. on the magnitude of the strain $\delta_y$ as suggested by figure~\ref{fig3}(b). This figure exhibits the average mass of wire $m(r)$ within the type of circular blobs shown in the upper left inset of the figure, as a function of the radial variable $r$ defined at the interior of each blob. The power-law fit in this figure gives $m \sim r^{1.9 \pm 0.1}$, which is an indicative of the nearly two-dimensional character of the distribution of mass within the blobs after averaging in 24 equivalent blobs.

\begin{figure*}[!]

\centering \includegraphics{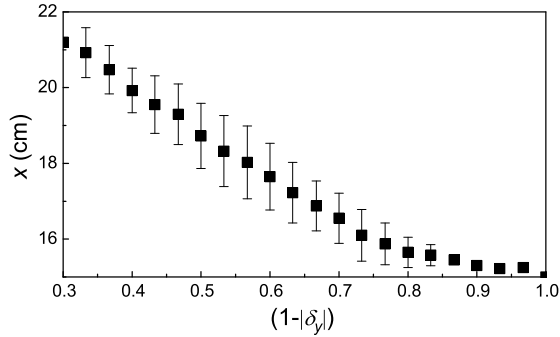}

\caption{\label{fig4}The average equatorial diameter $x$ indicated in figure~\ref{fig2}(b) plus the statistical fluctuations as a function of $1-|\delta_y|$. See text for details.}

\end{figure*}

\indent If the 2D-CW is compressed along an axis, its initial approximate circular symmetry shown in figure~\ref{fig1}(b) is broken. The process of transference of mass of wire perpendicular to the direction $y$ of the strain is described in figure~\ref{fig4}. This plot shows the average equatorial diameter $x$ indicated in figure~\ref{fig2}(b) plus the statistical fluctuations as a function of $1 - |\delta_y|$, in a wide interval, from $|\delta_y| = 0.033$ to $|\delta_y| = 0.69$ in steps of $\Delta|\delta_y| = 0.033$. Figure~\ref{fig4} can be understood with the following simple blob argument valid for both axially deformed crumpled surfaces~\cite{3}, and crumpled wires~\cite{15} in three dimensions. One assume that the crumpled wires at high strain behave as a collection of $n$ fractal blobs of diameter $y$, and mass $m \sim y^D$. Inside each blob the effects of the boundaries are weak, and successive blobs act as hard discs and pack into a regular 1D array of $n$ blobs satisfying $n = M/m = x/y$, and then $x = (M/m)y = (2R/y)^D y = (2R)^D y^{(1-D)} = 2R(1 - |\delta_y|)^{-\xi}$ , with $\xi = \xi(D) = D - 1$, if use is made of the definition of $\delta_y$ given in the beginning of section~\ref{sec2}. The last equation predicts for a system of mass fractal dimension $D = 1.9 \pm 0.1$, the scaling law $x \sim (1 - |\delta_y|)^{-(0.9 \pm 0.1)}$, which is, in fact, quite close to the dependence exhibited in figure~\ref{fig4}.

\begin{figure*}[!]

\centering \includegraphics{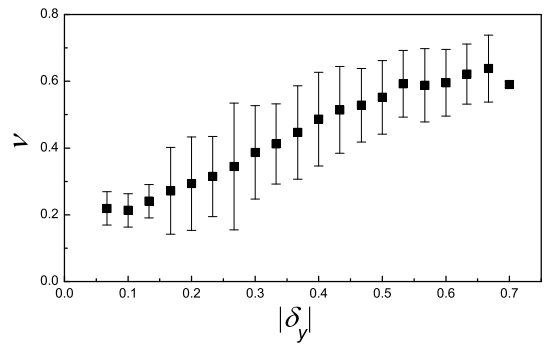}

\caption{\label{fig5}Experimental data for the average Poisson ratio $\nu$ (plus statistical fluctuations) as a function of the strain $|\delta_y|$.}

\end{figure*}

\indent The Poisson ratio~\cite{27} is defined as the ratio of the transversal and the axial strains introduced in section~\ref{sec2}, $\nu = - \delta_x/\delta_y = (x - 2R)/(2R - y)$.  The negative sign is included in the expression so that it will always be positive, since $\delta_x$  and $\delta_y$ will always be of opposite sign. For very small strain, the Poisson ratio is a constant belonging to the intervals (-1, 0.5), for 3D systems, and (-1,1), for 2D systems~\cite{28}. As a matter of illustration, in the linear elastic limit, the interval $0.25 < \nu < 0.35$, includes the Poisson ratio of most alloys, irons, and composites. Commercially pure gold has $\nu = 0.42$; lead and tin, both taking part of the alloy used in the wire (figures~\ref{fig1}(b) and~(c)), have in the same elastic limit Poisson ratios of 0.44, and 0.36, respectively. Differently, for high strain, the Poisson ratio is a function of the strain, as exemplified below for 2D-CW. As there is a coupling between both strains, $\delta_x$ and $\delta_y$, one can plot $\nu$ as a function for instance of $|\delta_y|$. This is done in figure~\ref{fig5}, which shows the average experimental Poisson ratio $\nu$ plus statistical fluctuations. The fluctuations bars in figure~\ref{fig5} represent the full interval of variability of $\nu$ for all samples used in the experiment. Although these fluctuations are not too small, because they reflect the high level of heterogeneity of the structures, the average values of $\nu$ are quite well described by a linear increasing function of $|\delta_y|$, from $|\delta_y| \approx 0.1$ to $|\delta_y| \approx 0.66$.

\begin{figure*}[!]

\centering {\includegraphics{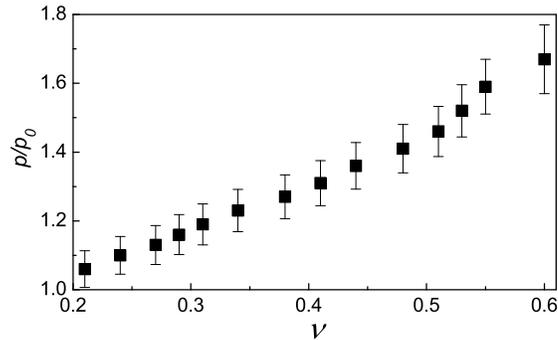}}

\caption{\label{fig6}Normalized packing density $p(\nu)/p_0$ as a function of the Poisson ratio $\nu$, where $p_0$ is the packing density at $\delta_y = 0$.}

\end{figure*}

\indent Finally, it was investigated the dependence of the packing fraction of the 2D-CW with the vertical strain. This quantity changes from $p_0 = 0.31 \pm 0.03$, for $\delta_y = 0$ to $p = 0.52 \pm 0.03$, for $|\delta_y| = 0.65 \pm 0.05$; i.e. it increases in $67\%$ in this interval of strain. It is interesting to notice that this interval of variability of packing fraction for 2D-CW when the axial strain varies in a wide interval is close to the variability observed in the critical probability $p_c$ in 2D bond percolation for the most studied lattices ($p_c = 0.34$; $0.49$; and $0.64$ for triangular; square; and honeycomb lattices)~\cite{29}. The value $p = 0.52$ obtained for the largest strains in our experiment, represents only $62\%$ of the maximum packing fraction for the random close packing of discs in 2D or $57\%$ of the packing fraction $\pi/2 \sqrt{3} = 0.9068 \ldots$ of the triangular close packing of discs, the densest possible arrangement of equal discs. Using the data in figure~\ref{fig5} and the experimental dependence $p(|\delta_y|)$ we obtained the dependence of the packing fraction with the Poisson ratio. Thus, the packing fraction $p$ is an increasing function of $\nu$ as shown in figure~\ref{fig6}. The Poisson ratio and the packing fraction $p$ or the porosity $1-p$ are basic properties of materials in general. For disordered materials as the crumpled wires discussed here, the relationship between these quantities is of particular recent interest~\cite{28,30}. In fact, the plot in figure~\ref{fig6} obtained for our 2D structures of loops and lamellae illustrated in figures~\ref{fig1}(b) and~\ref{fig1}(c) is reminiscent of a similar plot shown in~\cite{30} for a planar ensemble of cellular structures in soft media acting as force dipoles.

\section{\label{sec4} Summary and Conclusions}

\indent Besides its intrinsic interest, low-dimensional structures of crumpled wires could be related to random network motifs of atoms or molecules adsorbed on solid surfaces~\cite{24}, to the study of DNA packing in chromosomes and viral capsids~\cite{21}, and to describe the statistical aspects of meandering nanotubes in planar confining geometries~\cite{25}, among others. Here, it has been studied experimentally the geometric changes observed in packings of crumpled wires of circular shape when such structures are confined in a 2D cell and an axial compressive strain is applied. In particular, one have observed scaling laws connecting variables of physical interest as (i) mass and size, (ii) relative lateral expansion and axial strain, (iii) Poisson function and strain, and (iv) volume fraction as a function of the Poisson ratio, among others. Surprisingly, critical exponents as fractal dimensions were found to be independent of the strain in a very large interval of variability.

\section*{Acknowledgments}

This work was supported in part by Conselho Nacional de Desenvolvimento Cient\'{\i}fico e Tecnol\'ogico, and Programa
de N\'ucleos de Excel\^encia (Brazilian Government Agencies). C.C.D. acknowledges a postdoctoral fellowship from CNPq.

\end{document}